

\input harvmac.tex
\noblackbox
\sequentialequations
\lref\GSW{M. Green, J. Schwarz, and E. Witten, {\it Superstring
Theory}, Vol. 2, Cambridge University Press (1987) 357.}
\lref\AHW{I. Affleck, J. Harvey, and E. Witten, ``Instantons
and (Super-)Symmetry Breaking in (2+1) Dimensions,'' {\it Nucl. Phys.}
{\bf B206} (1982) 413.}
\lref\atish{A. Dabholkar, ``The Ten-Dimensional Heterotic String
as a Soliton,'' hep-th/9506160.}
\lref\hull{C. Hull, ``String-string Duality in Ten-Dimensions,''
hep-th/9506194.}
\lref\SS{J. Schwarz and A. Sen, ``The Type IIA Dual of the Six-Dimensional
CHL Compactification,'' hep-th/9507027.}
\lref\SchSen{J. Schwarz and A. Sen, ``Duality Symmetries of 4d Heterotic
Strings,'' Phys. Lett. {\bf B312} (1993) 105.}
\lref\HullTown{C. Hull and P. Townsend, ``Unity of Superstring Dualities,''
hep-th/9410167.}
\lref\BBS{K. Becker, M. Becker, and A. Strominger, ``Three-Dimensional
Supergravity and the Cosmological Constant,'' {\it Phys. Rev.} {\bf D51}
(1995) 6603.}
\lref\Ed{E. Witten, ``String Theory Dynamics in Various Dimensions,''
hep-th/9503124.}
\lref\desjac{S. Deser and R. Jackiw, ``String Sources in 2+1-Dimensional
Gravity,'' {\it Ann. Phys.} {\bf 192} (1989) 352.}
\lref\edthree{E. Witten, ``Is Supersymmetry Really Broken?,'' {\it Int.
J. Mod. Phys.} {\bf A10} (1995) 1247, hep-th/9409111 \semi
E. Witten, ``Strong Coupling and the Cosmological Constant,'' hep-th/9506101.}
\lref\joyceG{D. Joyce, ``Compact Riemannian 7-Manifolds with
Holonomy $G_2$: II'', to appear in {\it J. Diff. Geom.} }
\lref\joyceS{D. Joyce, ``Compact Riemannian Manifolds with
Exceptional Holonomy,'' Oxford preprint.}
\lref\samvaf{S. Shatashvili and C. Vafa, ``Superstrings and Manifolds
of Exceptional Holonomy,'' hep-th/9407025.}
\lref\chs{C. Callan, J. Harvey, and A. Strominger, ``Supersymmetric
String Solitons,'' hep-th/9112030.}
\lref\VW{C. Vafa and E. Witten, ``Dual String Pairs with N=1 and N=2
Supersymmetry in Four Dimensions,'' HUPT-95/A023, IASSNS-HEP-95-58,
hep-th/9507050.}
\lref\Lowe{J. Harvey, D. Lowe, and A. Strominger, ``N=1 String Duality,''
hep-th/9507168.}
\lref\kv{S. Kachru and C. Vafa, ``Exact Results for N=2 Compactifications
of Heterotic Strings,'' hep-th/9505105, to appear in Nucl. Phys. {\bf B}.}
\lref\fhsv{S. Ferrara, J. Harvey, A. Strominger, and C. Vafa,
``Second Quantized Mirror Symmetry,'' hep-th/9505162.}
\lref\anom{C. Vafa and E. Witten, ``A One-loop Test of String
Duality,'' hep-th/9505053.}
\lref\orb{J. Bagger, C. Callan, and J. Harvey, ``Cosmic Strings
as Orbifolds,'' {\it Nucl. Phys.} {\bf B278} (1986) 550.}
\lref\scs{B. Greene, A. Shapere, C. Vafa, and S-T Yau, ``Stringy
Cosmic Strings and Noncompact Calabi-Yau Manifolds,'' {\it Nucl. Phys.}
{\bf B337} (1990) 1.}
\lref\sen{A. Sen, ``String-string Duality Conjecture in Six-Dimensions
and Charged Solitonic Strings,'' hep-th/9504027.}
\lref\harvstrom{J. Harvey and A. Strominger, ``The Heterotic String
is a Soliton,'' hep-th/9504047.}
\lref\hpol{J. Hughes and J. Polchinski, ``Partially Broken Global Supersymmetry
and the Superstring,'' {\it Nucl. Phys.} {\bf B278} (1986) 147.}

\Title{\vbox{\hbox{HUTP-95/A029}\hbox{PUPT-1556}\hbox{\tt hep-th/9508096}}}
{\vbox{\centerline{Nonsupersymmetric String Solitons}
        }}
\centerline{Shamit Kachru\footnote{$^*$}
{kachru@string.harvard.edu}}
\medskip\centerline{Lyman Laboratory of Physics}
\centerline{Harvard University}\centerline{Cambridge, MA 02138}
\bigskip\centerline{Eva Silverstein\footnote{$^\dagger$}
{silver@puhep1.princeton.edu}}
\medskip\centerline{Joseph Henry Laboratories}
\centerline{Jadwin Hall}
\centerline{Princeton University}\centerline{Princeton, NJ 08544}
\vskip .3in

We begin a search for nonsupersymmetric/supersymmetric dual string
pairs by constructing candidate critical
nonsupersymmetric strings as solitons in
supersymmetric string theories.
Using orbifold techniques, one can construct cosmic
string solutions which lie in supersymmetric vacua but which do
not fall in supermultiplets.  We discuss two three-dimensional
examples in detail.
The effective worldsheet actions for the soliton strings
have (0,2) and (1,1) supersymmetry
and the correct numbers of massless bosons and fermions to be critical
heterotic and type II strings, respectively.

\Date{August 1995} 

\newsec{Introduction}

It is a well-known fact in string theory that worldsheet
supersymmetry alone does not imply spacetime supersymmetry.
In this paper, we attempt to exploit this fact by constructing
nonsupersymmetric strings as solitons in supersymmetric string
vacua.  In six dimensions, a solitonic solution of type IIA string theory
compactified on $K3$ has been found with an effective
worldsheet action identical to critical heterotic string theory
compactified on $T^4$ \sen\harvstrom\ (a similar construction has been given
of the
10d $SO(32)$ heterotic string as a soliton in the Type I theory \atish\hull).
In this construction, the
spacetime supersymmetry left unbroken by the soliton leads to worldsheet
supersymmetry on the cosmic string, while the normalizable broken supercharges
create the superpartners of the soliton \hpol\harvstrom.

Our purpose here is to present a similar (though much weaker) construction of
critical string solitons in supersymmetric theories, with unbroken
spacetime supersymmetry leading to worldsheet supersymmetry,
but without the broken supersymmetry generators which would have produced
target space supersymmetry.  The absence of broken supercharges for
string solitons which nevertheless have finite action is a
feature of the stringy
cosmic string and orbifold cosmic string constructions of \scs\ and
\orb, and we use essentially the methods of \orb\ to construct
our examples (a field theory example of this kind of
phenomenon based on \edthree\ was studied in \BBS).
In these constructions, the spacetime and internal
degrees of freedom are tied together in such a way as to produce
a noncompact manifold admitting some number of covariantly
constant spinors.  In order to have well-defined broken
supersymmetries, there must be additional spinor fields
which are not covariantly
constant but which asymptote to covariantly constant spinors far
away from the string core.  Such spinors do not exist in
the orbifold and noncompact Calabi-Yau constructions.
Still, as explained by Bagger, Callan, and Harvey for the orbifold
examples \orb, far away from
the string one recovers the full spectrum of the original vacuum.
We will review this as it becomes relevant for us.

Recent work has demonstrated
that the string-string duality relating 6d N=2 heterotic and type IIA vacua
\HullTown\Ed\SS\ (which implies the S-duality of 4d N=4 string
theories \SchSen)
extends also to
4d $N=2$ examples \kv\fhsv\ and to some 4d
$N=1$ theories in which no superpotential is generated \VW\Lowe.
In some of the 4d $N=2$ examples, the absence of neutral couplings
between vector and
hypermultiplets ensures that the
duality allows one to compute quantum properties of one
vacuum from the purely classical physics of its dual.
In $N<2$ theories, one might hope that there is a duality in which
the nonperturbative properties of one vacuum
are manifest in the perturbative physics of another.
The consequences
of such a duality for dynamical N=1 theories would be extremely interesting.
For four-dimensional N=1 models which break supersymmetry, as
well as for three-dimensional supersymmetric models \edthree,
such a structure would imply that we should expect
{\it non-supersymmetric} duals for supersymmetric theories.
This motivates us to search for candidate nonsupersymmetric
dual strings in supersymmetric string vacua.

The construction presented here is preliminary and limited
in several important respects.
In the six-dimensional construction, the BPS mass formula for
six-dimensional $N=2$ supersymmetry
is sufficient to ensure that such a soliton
becomes light at strong coupling and therefore provides a dual description
of the original theory.  In our case, as in any situation with
4d $N\le 2$ supersymmetry, the question is more delicate.  It might
require solving the theory at low energy to show convincingly that a soliton
constructed at weak coupling becomes light anywhere.
We also have little control over the effective worldsheet theory we obtain
on the soliton beyond counting its modes (requiring a critical
number) and counting the supersymmetries.
However we believe that it is worth searching for
circumstantial evidence for nonsupersymmetric/supersymmetric
dual pairs through string
solitons, since it is
difficult to see what to compare in more direct approaches.

\newsec{General Constraints}

In which dimensions should we expect to find cosmic string
solutions which do not fall into supermultiplets?
Orbifold cosmic strings are constructed
using a discrete group which acts simultaneously on the internal
compactification and
on the transverse spacetime dimensions \orb, i.e. by
taking orbifolds of $T^{8}$ or $T^{8-d}\times R^{d}$.  This prescription
always has the property that no broken supercharges are defined,
as discussed above.  However, it only
produces a cosmic string geometry in at most four dimensions.
So fortunately this construction can only work in four or
fewer dimensions, where it is possible to break supersymmetry.

Starting from
heterotic or Type I string theory, one could obtain for
example (0,2) supersymmetry on the string worldsheet
by constructing an eight-dimensional orbifold
with holonomy $SU(4)$, or (0,1) supersymmetry
from an eight-fold of holonomy ${\rm Spin(7)}$
(in Table 1 we assemble a list of the
surviving supersymmetries for the various string theories and holonomy
groups).  However a stable solution is not obtained
with just any eight-dimensional
orbifold which preserves spacetime supersymmetry \anom.
The eight-dimensional effective action contains the Green-Schwarz
term $\int B X_8$ where $X_8$ is an eight-dimensional differential
form involving the Riemann tensor and the gauge field strengths \GSW.
A term like this arises in the Heterotic, Type I,
and Type IIA theories, destabilizing solitons for which it
does not vanish.  In particular it does not vanish for generic
eight-manifolds with $SU(4)$ or ${\rm {Spin(7)}}$ holonomy.
There is another potentially destabilizing effect, discussed
in \orb --the one-loop contribution to the two-dimensional
vacuum energy, proportional to the difference between the number
of massless bosons and fermions.
We will consider one example
which is unstable at one loop, and one
in which neither of these effects arises.  When such terms arise, one plausible
possibility is that the configuration relaxes to a stable one
with the same worldsheet structure as the unstable one \orb.

\def\tablerule{\omit&\multispan{8}{\tabskip=0pt\hrulefill}&\cr}
\def\tablepad{\omit&height3pt&&&&&&&&&\cr}
$$\vbox{\offinterlineskip\tabskip=0pt\halign{\hskip 1.0in
\strut$#$\quad&\vrule#&\quad\hfil $#$ \hfil\quad &\vrule #&\quad \hfil $#$
\hfil \quad&\vrule #& \quad $#$ \hfil\ &\vrule #&\quad $#$ \hfil\
&\vrule#&\quad $#$\cr
&\omit&\hbox{String Theory}&\omit&\hbox{SU(4)}
&\omit&\hbox{~$\rm {G_{2}}$}&\omit&\hbox{Spin(7)
}&\omit&\cr
\tablerule\tablepad
&&{\rm Heterotic/Type~ I}&&(0,2)&&(1,1)&&~(0,1)&&\cr
\tablerule\tablepad
&&{\rm Type ~IIA}&&(2,2)&&(2,2)&&~(1,1)&&\cr
\tablerule\tablepad
&&{\rm Type ~IIB}&&(0,4)&&(2,2)&&~(0,2)&&\cr
\tablerule
\noalign{\bigskip}
\noalign{\narrower\noindent{\bf Table 1:} Worldsheet SUSY of soliton strings
coming from string compactifications on manifolds with various holonomy
groups.}
 }}$$

In order to guarantee the absence of supersymmetry in the dual theory, one
can take the soliton worldsheet to have only (0,1) or (1,1) supersymmetry.
The critical (0,2) and (2,2) cases are more delicate.  We will consider
a (0,2) example
in section 3 and a geometrically
puzzling (1,1) example in
section 4.  Because of the
absence of broken supercharges and the corresponding lack of superpartners
for the cosmic strings, one would expect the (0,2) and (2,2) examples also to
fail to have target space supersymmetry.
That would imply that the U(1) R-charges are not integrally
quantized on such a soliton worldsheet.  We do not yet have
enough control over the worldsheet theory to check this for
the example in section 3.

\newsec{A (0,2) String in a Type IIB Vacuum}

In this section we discuss an example of a (0,2) string
soliton in a $3d$ Type IIB vacuum (with the string wrapped
around one of the internal seven dimensions).  We will find 8 nonchiral
bosonic modes (along with their (0,2) superpartners)
as well as 16 chiral bosons on the soliton worldsheet.
Begin with $Z_2\times Z_2$
generators acting on the spacetime coordinates $x^1,\dots,x^{9}$ as
follows:

\eqn\A{A:x\rightarrow (-x^1,-x^2,-x^3,-x^4,x^5,x^6,x^7,x^8,x^9)}
\eqn\C{C:x\rightarrow (-x^1,{1\over 2}-x^2,x^3,x^4,{1\over 2}-x^5,
{1\over 2}-x^6,x^7,x^8,x^9)}

We take $x_{7,8}$ to be $\it noncompact$ -- they will play the role
of the conical geometry transverse to the string momentarily.
If we compactify $x^9$, the direction along which the string lies,
we have so far a $3d$ $N=4$ vacuum.  We will see later
that the compactification of $x^9$
is necessary to obtain a critical number of worldsheet modes.
To obtain a string, we include another $Z_2$ as follows:

\eqn\B{B:x\rightarrow (x^1,x^2,x^3,x^4,-x^5,-x^6,-x^7,-x^8,x^9)}

These three generators give an eightfold of $SU(4)$ holonomy considered
by Joyce \joyceS.  So far the worldsheet supersymmetry is
(0,4).  The effective worldsheet theory is populated by
the states tied to the string, namely the states from the $B$ and
$AB$ twisted sectors (there are no massless states from the $BC$ sector).
We can count these states using dimensional reduction
on the blown up orbifold described by Joyce \joyceS.
The (NS, NS) sector contributes the metric and antisymmetric tensor
fields (as well as the dilaton, which does not lead to new states
under dimensional reduction).
As discussed in \joyceS, the $B$ sector blow up introduces
three extra metric deformations and one additional two-form per fixed torus.
$B$ has four fixed tori to begin with but $C$ exchanges them in two pairs
of two.  So we have a total of eight modes on the string from the
(NS, NS) $B$ sector.  These modes have no potential on the
worldsheet since they are (2,2) moduli of the original string theory.

The (R,R) sector contributes another two-index
antisymmetric tensor field and a self-dual four-form.
The vacuum in the (R,R) $B$ sector is an $SO(5,1)$ spinor from the
fermions corresponding to $x^1,\dots, x^4,x^9,x^{10}$. The GSO and
BRST projections
reduce this to states which have the same $SO(4)$ and $SO(1,1)$ chiralities
on left and right, where the $SO(4)$ acts on $x^1,\dots,x^4$ and
the $SO(1,1)$ on $x^9$ and $x^{10}$.
These are invariant under the $A$ projection, which acts the same
on the left and right $SO(4)$ spinors.
These massless states can be projected out by another $Z_2$ generated by $D$:
$D$ acts similarly to $A$ except that it
acts oppositely on the
left and right-moving SO(4) spinors.
In order to avoid massless states in the $D$ twisted sectors, we include
a shift on $x^9$.  We also include a shift on $x^1$ in addition
to the minus sign on that coordinate which will give us the
desired degeneracy in the $AB$ sector.
$D$ projects out the massless states from the (R,R) and (R, NS)
$B$ sectors, as well as projecting out the gravitinos coming from the
left-movers.  This projection thus
leaves us with the 8 (NS, NS) $B$ sector bosons
counted above while reducing the vacuum supersymmetry to $3d$ $N=2$ and
the soliton worldsheet supersymmetry to (0,2).

What about the chiral modes on the heterotic soliton worldsheet?
These come from the (R,R) $AB$ sector.  Here the vacuum is unique, and
before the $C$ and $D$ identifications there are $2^6=64$ fixed points.
Because
of their shifts, $C$ and
$D$ reduce
this by a factor of 4, leaving 16 bosons from the $AB$ sector.
That these modes are chiral follows from the fact that the
$AB$ NS sectors have positive vacuum energy ($E_{AB}^{NS}=1/2$).
So there are no supersymmetric partners of the 16 (R,R) bosons
from the $AB$ sector, indicating that they are on the nonsupersymmetric
side of the candidate critical heterotic soliton string worldsheet.
These modes can also be understood from dimensional reduction: as
explained in \joyceS, blowing up an $AB$ fixed point introduces
one additional self-dual four-form, leaving all the other betti
numbers fixed.  This leads to one additional scalar per fixed point
upon dimensional reduction of Type IIB string theory.

As discussed in \orb, away from the string we recover the
original vacuum spectrum.  Given a vertex operator $V$ which
is antiinvariant under $B$, there is an associated invariant vertex operator
$\tilde V=(e^{i(p_7x_7+p_8x_8)}-e^{-i(p_7x_7+p_8x_8)})V$.  Because
$x^7$ and $x^8$ are noncompact, the momenta $p_7$ and $p_8$ can
chosen to yield a massless state ($p^2$=0).
The vacuum in this theory perturbatively has only abelian gauge symmetry
with no charged matter,
and therefore uninteresting dynamics.  However, as discussed in
\edthree, in three dimensions massive particles do not fall into
supermultiplets in any case.  Therefore our candidate duality is possible.

\newsec{A (1,1) String in a Heterotic Vacuum}

We are now going to consider a candidate critical type II string, obtained
from compactification of the heterotic string
on a manifold of $G_{2}$ holonomy.
While it is impossible to find cosmic strings in $\it four$ spacetime
dimensions by compactification on an orbifold of $G_{2}$ holonomy, we
will now briefly explain that we can obtain three dimensional cosmic strings
through compactification on $X\times S^{1}$ with $X$ a (noncompact)
manifold of $G_{2}$ holonomy.  It is important to make a distinction between
this construction and the previous one, in which the string wound around
a circle in the internal space but the transverse geometry consisted of
a two-dimensional cone.   In the present construction, the string will be
intrinsically 2+1 dimensional with the transverse space being one dimensional.

We must first understand the classical geometry of a string in 2+1 dimensions.
This was studied by Deser and Jackiw \desjac.  They found that a delta-function
localized closed string with tension produces a geometry consisting of a
capped half-infinite cylinder, with the string at the boundary of the cap.
With tension, the mass is fixed at $M=G^{-1}$ (where $G$ is Newton's
constant), yielding the $2\pi$ deficit angle.

The half-cylinder part of this spacetime, including the localized cosmic
string, can be produced by the orbifold technique.  Consider a compactification
down to three dimensions on a Calabi-Yau manifold $K$ times a circle,
accounting for dimensions $x^{0}$ through $x^{6}$.  Orbifold by a $Z_{2}$
which maps $x^{7}\rightarrow -x^{7}$ and simultaneously acts by an involution
on $K$ such that the resulting eight-fold has holonomy $G_{2}$ (such
examples were introduced by Joyce \joyceG\ and discussed in the context
of conformal field theory in \samvaf).  If we compactify $x^{8}$, we
obtain a half-cylinder geometry.  Furthermore, we have localized stress-energy
at $x^{7}=0$ from the internal components of the metric, which appears to
the three-dimensional observer as gradient energy from the moduli.  So we
have recovered the part of the Deser/Jackiw solution external to and
including the string.

Classically, we could now patch on the flat geometry internal to the string
(the cap of the half-cylinder), completing the geometry to a solution of
the classical equations of motion.  Quantum mechanically, the situation
is less clear: In nonrelativistic quantum mechanics, propagation through
a delta-function wall occurs perturbatively in Planck's constant, and it
is hard to see how this would be mirrored in our orbifold conformal field
theory describing the half-cylinder.  But it is not completely clear that
this is the behavior in our case.  If in our situation the tunneling
through the string is nonperturbative, then the patching construction is
valid, and we have a closed string in three dimensions in line with the
solution of \desjac.  Without the patching, we can still interpret our
construction as a closed string, but in a non-simply connected spacetime.
In either case, we obtain a closed string instead of the infinitely extended
cosmic strings which occur in the higher-dimensional dual string soliton
constructions of \sen\harvstrom\atish\hull.

The example presented here is based upon modifications of one of the $G_{2}$
orbifolds constructed
by Joyce \joyceG, and we use some of his notation.
We begin with the $SO(32)$ heterotic string compactified on
$S^1 \times K$ where K is
a Calabi-Yau manifold realized as a toroidal orbifold.
Let $x^0$ be the coordinate along the circle and
$x^1,\dots x^6$ the coordinates on the $T^6$.
The group elements producing K act on $T^6$ as follows:
\eqn\a{\alpha : x\rightarrow (-x^1,-x^2,-x^3,-x^4, x^5,x^6)}
\eqn\b{\beta : x\rightarrow
(-x^1, {1\over 2}-x^2,x^3,x^4,-x^5,-x^6)}
\eqn\sI{s_1: x\rightarrow ({1\over 2}+x^1,{1\over 2}+x^2,x^3,x^4,x^5,x^6)}
\eqn\sII{s_2: x\rightarrow (x^1,x^2,{1\over 2}+x^3,{1\over 2}+x^4,x^5,x^6)}
\eqn\sIII{s_3: x\rightarrow (x^1,x^2,x^3,x^4,{1\over 2}+x^5,{1\over 2}+x^6)}
We use the standard embedding of these group elements
into the gauge degrees of freedom.
So far we have a three-dimensional $N=2$ compactification on
$S^1\times K$.

To produce the cosmic string, we include another
group element $\gamma$, which acts on one noncompact transverse dimension $x^7$
as well as on $K$.
\eqn\gam{\gamma: x\rightarrow ({1\over 2}-x^1,x^2,-x^3,x^4,-x^5,x^6,-x^7)}
We use the standard embedding for $\gamma$ as well.
The dimensions $x^8$ and $x^9$ form the cosmic string worldsheet.
The orbifold produced by these group elements (which can
be blown up to a manifold of $G_2$ holonomy \joyceG) leaves
(1,1) supersymmetry on the string world sheet.  Our construction
will leave 8 bosonic zero modes on the string world sheet,
consistent with the worldsheet action for a  critical type II string.
As explained above, we will compactify $x^8$ (and possibly insert
a flat disc at $x^7=0$) to obtain a closed string solution in 2+1 dimensions.
The only group elements with fixed
points (and therefore candidate massless states) are
$\alpha$, $\beta$, and $\gamma$.

The modes tied to the string
come from the $\gamma$ twisted sector \orb.
We will later introduce one more group element, $g$, to reduce the degeneracy
of $\gamma$-twisted states to a critical number.  As discussed
in \orb, despite the $\gamma$ projection we recover the full
vacuum spectrum away from the string in the untwisted, $\alpha$,
and $\beta$ sectors.  This is because the direction $x^7$ is noncompact.
Given a vertex operator $V$ with $\gamma$ eigenvalue -1, we
can construct another vertex operator
$\tilde V\equiv V(\exp{(ip_7x^7)}-\exp{(-ip_7x^7)})$ for $x^7\ne 0$ which is
$\gamma$-invariant.  Because $p_7$ is not quantized, we can arrange
that $p^2=p_7^2+p_8^2-p_9^2=0$ for a massless state.

To understand the string worldsheet action, we are interested
in the states from the $\gamma$ sector.  Denote the
left-moving real fermions $\eta^I$ where $I=1,\dots,32$ and
the right-moving real ferions $\psi^i$.  For the bosons
(right-moving Neveu-Schwarz sector) the right and
left-moving vacuum energies in the $\gamma$ sector are
\eqn\enR{E_R^{NS} = 0}
\eqn\enL{E_L^A = -1/2}
where $E_L^A$ is the left-moving vacuum energy for the sector
in which the $\eta^I$ are naturally antiperiodic.  The
periodic sector has positive vacuum energy and yields no massless states.
Both left and right have 4 real fermion 0-modes in this sector.  Group
them into complex fermions $\lambda_1=\eta_1+i\eta_3$ and
$\lambda_2=\eta_5+i\eta_7$ on the left and
$\chi_1=\psi_1+i\psi_3$ and $\chi_2=\psi_5+i\psi_7$  on the right.
There are also 4 twisted spactime bosons
yielding creation operators $\alpha_{-1/2}^{1,3,5,7}$.

Before the $\alpha$ and $\beta$ projections there are states of
the following form

\eqn\stsI{S_+  \alpha_{-1/2}^{1,3,5,7} |-1/2 \rangle \otimes \tilde S_- |0
\rangle}

and

\eqn\stsII{S_- \eta_{-1/2}^I |-1/2 \rangle \otimes \tilde S_- |0 \rangle}
where $S_\pm$ is a $\pm$-chirality spinor from the $\lambda_{1,2}$
0-modes and $\tilde S_-$ is a negative-chirality spinor from
the $\chi_{1,2}$ 0-modes.
The $S_+$ spinor is obtained by acting with $\bar\lambda_1$ or
$\bar\lambda_2$ on
the vacuum $|-1/2\rangle$ killed by $\lambda_{1,2}$; the
$S_-$ spinor has components $|-1/2\rangle$ and
$\bar\lambda_1\bar\lambda_2|-1/2\rangle$.
The I index runs over the 28 $\eta$s which
are antiperiodic in the gamma sector.  It is easy to see that these
states are all $\gamma$-invariant.

It follows from \a\ and \b\ that $\alpha$ acts with -1 on $\bar\lambda_1$ and
$\bar\chi_1$ and
with +1 on $\bar\lambda_2$ and $\bar\chi_2$, while $\beta$ acts by
anticonjugating all the $\lambda$s and
$\chi$s.  In order to form $\alpha$ and $\beta$-invariant states
we first need to find combinations of the left and right-moving
spinor components that are eigenstates of both $\alpha$ and
$\beta$.  For states of the form \stsI,
these are provided by the following combinations:
\eqn\V{V_\pm \equiv \bar\lambda_1 \otimes 1
\pm \bar\lambda_2\otimes\bar\chi_1\bar\chi_2}
which has $\alpha=-1$ and $\beta=\mp 1$, and
\eqn\W{W_\pm \equiv \bar\lambda_1\otimes \bar\chi_1\bar\chi_2
\pm \bar\lambda_2 \otimes 1}
which has $\alpha=+1$ and $\beta=\mp 1$.
Using these we find invariant states
$\alpha_{-1/2}^1 V_+$, $\alpha_{-1/2}^3 V_-$,  $\alpha_{-1/2}^7 W_-$, and
$\alpha_{-1/2}^5 W_+$.
This gives four states on the string so far. (Recall from the shifts in
\sI, \sII, and \sIII\ that there is only one $\gamma$ fixed point.)

What about states of the form \stsII?
First of all, let's project a lot of them out by the following extra $Z_2$
generated by $g$.  Let $\lambda_3=\eta_2+i\eta_4$ and
$\lambda_4=\eta_6+i\eta_8$.
Have $g$ act with -1 on $S_-$ and on $\lambda_3$ and $\lambda_4$, and
also a $Z_2$ shift on $x^0$ to avoid massless states in any $g$-twisted sector.
The action on $\lambda_3$ and $\lambda_4$ is included for level matching.
Before the $\alpha$ and $\beta$ projections, this leaves states of the form

\eqn\stsIItrunc{S_- \lambda_{-1/2}^{3,\bar 3,4,\bar 4} |-1/2 \rangle
\otimes \tilde S_- |0 \rangle}
By making the analogous combinations as described above for
the \stsI\ states, we find another 4 states on the string from these.
We have thus a total of 8 massless bosonic modes (along with their
(1,1) fermionic superpartners) on our closed string world sheet.

It turns out that we can learn more than just the number of massless modes.
We must further check that our purported zero modes are
gauge-neutral and have a flat potential on the worldsheet.
The gauge group in our three-dimensional vacuum given
by the $\alpha$, $\beta$, $s_{1,2,3}$, and $g$ orbifold
is
\eqn\gauge{G=U(1)^2\times SO(24)\times SO(2)^4.}
The first $U(1)^2$ factor comes from the dimensional reduction
of the $4d$ metric and antisymmetric tensor field on the circle $x^0$.
Our $\gamma$-sector worldsheet states \stsI\ and \stsIItrunc\ are neutral under
the $SO(24)$ factor here.
The $SO(2)^4$ gauge generators arise from the worldsheet currents
$\eta^1\eta^2$, $\eta^3\eta^4$, $\eta^5\eta^6$, and $\eta^7\eta^8$.
Since these operators are odd under $\gamma$, the corresponding
gauge boson vertex operators are constructed using
the momentum $p_7$ and only exist away from the string.
We will now demonstrate that our worldsheet modes are flat.
If we leave the
$\alpha$ and $\beta$ fixed points untouched (i.e. not
turning on the blowing up modes from these sectors)
this follows from the fact that the
untwisted and $\gamma$-sector scalars form a subset of the
scalar spectrum of heterotic string theory compactified on
$K3$.  That is, $\gamma$ and $g$ alone produce
a compactification on $K3$.  The states from the untwisted and
$\gamma$ sectors which survive the further projections onto
$\alpha$, $\beta$, and $s_{1,2,3}$-invariant states
form a subset of the $K3$ spectrum.  These states inherit
their correlation functions from the $Z_2$ orbifold
$K3$ given by $\gamma$ and $g$, which is governed
by $N=2$ supersymmetry in four
dimensions.  The scalar potential arises from a highly constrained
superpotential and from $U(1)$ $D$-terms, in $N=1$ language.
The superpotential arises for gauge-charged hypermultiplets
and is a cubic coupling of the adjoint chiral multiplet $a$ with
the two chiral multiplets $\phi$ and $\tilde\phi$ in the hypermultiplet.
Our spectrum is not the full $K3$ spectrum but is truncated
by the $\alpha$ and $\beta$ projections.  This means in particular
that if $\phi$ is an invariant chiral multiplet then its
$N=2$ partner $\tilde\phi$
will not appear in the spectrum, as $\alpha$ removes one of the
two supersymmetries.  Therefore as long as $a = 0$,
there is no superpotential preventing us from giving the soliton
worldsheet modes
arbitrary VEVs.

We should also make sure that the worldsheet modes are not
obstructed by $D$-terms.  With
just the $\gamma$ and $g$ projections,
one has a gauge group $SO(4)^2\times SO(24)$ in the $K3$ theory
under discussion.  (Note that this spectrum of gauge bosons
has nothing to do with that of our vacuum theory generated
by $\alpha$, $\beta$, $s_{1,2,3}$ and $g$ given in \gauge; we are considering
it simply as a shortcut to determining the flat directions among
the untwisted and $\gamma$-sector states).
All our $\gamma$-sector scalars are
invariant under the $SO(24)$ factor.
Consider
a $U(1)^4$ subgroup of $SO(4)^2$ which is generated by
the worldsheet currents $\eta_1\eta_3$,
$\eta_5\eta_7$, $\eta_2\eta_4$, and $\eta_6\eta_8$.
Turning on our worldsheet modes leads to nonzero $D$-terms only
for these generators; this is also the case for the
VEVs we will turn on for vector representations in the untwisted sector.
Our set of worldsheet states \stsI\ and \stsIItrunc\ includes
states charged under each of these $U(1)$s.
Before the $\beta$ projection, in the untwisted sector
we find scalars of positive and negative charge under each $U(1)$.
For example we find the 16 states
\eqn\sone{(\eta^1\pm i\eta^3)_{-{1\over 2}}(\eta^6\pm i\eta^8)_{-{1\over 2}}
\vert{-1\rangle}
\otimes (\psi^1\pm i\psi^3)_{-{1\over 2}}\vert{-{1\over 2}\rangle}}
\eqn\stwo{(\eta^5\pm i\eta^7)_{-{1\over 2}}(\eta^2\pm i\eta^4)_{-{1\over 2}}
\vert{-1\rangle}
\otimes (\psi^1\pm i\psi^3)_{-{1\over 2}}\vert{-{1\over 2}\rangle}}
This allows us to
balance the $D$-terms while leaving
the $\gamma$-sector scalars free to vary. Now,
$\beta$ projects out
half of each $U(1)$ representation, preserving only the real
or imaginary part from the scalar component of each chiral multiplet.
The $D$-terms can still be made to vanish while allowing the
worldsheet scalars to vary freely by adjusting the VEVs of the untwisted
charged fields.
So we have 8 worldsheet scalars which appear to
lie on some real subspace of the moduli
space of $K3$.  It is not clear what the global structure of this
worldsheet sigma model is.  If it flows
to a nontrivial conformal field theory in the
infrared, then we have constructed a
critical nonsupersymmetric type II string as a soliton in a
three-dimensional $N=2$ compactification of heterotic string theory.

As mentioned in \S2, we
need to discuss the stability of our cosmic string when quantum corrections
are included.  In \orb\ the one-loop correction to the vacuum energy
for such orbifold cosmic strings, which should be interpreted as a
loop correction to the string tension (since it is independent of the size of
the transverse dimensions), was evaluated and found to be
proportional to the difference between the numbers of bosons and
fermions in two dimensions.  Because
our compactification leaves (1,1) supersymmetry in two dimensions,
this vanishes.\foot{In order to use the computation of \orb, we should
strictly speaking compactify $x_{7}$ as well.  This would introduce
another string soliton, but there would still be a critical number of
worldsheet modes on both string solitons.}
So at least at one loop we do not need to
worry about string decay through dilaton emission.
Also the part of $X_8$ depending only on the Riemann curvature
tensor vanishes for our solution, since there is one transverse
flat circle ($x^0$).  Thus our solution is stable to one loop.

Now that we have found a candidate critical nonsupersymmetric string soliton
in a $3d$ $N=2$ heterotic vacuum, we should discuss in a bit more detail the
physics of this vacuum.
As discussed above, the
standard embedding of $\alpha$, $\beta$, and $\gamma$ combined with
the extra $g$-projection reduces the gauge group to
$U(1)^2\times SO(24)\times SO(2)^4$.  All states in the $\alpha$ and
$\beta$ sectors are uncharged under the $SO(24)$ factor.
The states \stsI\ and \stsIItrunc\ are invariant under $SO(24)$.
The extra projection onto $g$-invariant
states removes the fundamental matter representation of $SO(24)$ from
the untwisted spectrum.
The four-dimensional vacuum obtained by taking
$x^0$ large consists of pure
$SO(24)$ $N=1$ supersymmetric Yang-Mills theory with some additional
neutral chiral multiplets, a theory which we do not
expect to spontaneously break SUSY.
This 4d theory is a limit in which our soliton string becomes a
two-brane whose properties are not well understood.
Compactifying $x^0$ to an $S^1$
produces scalars in the adjoint representation of $SO(24)$, yielding
a $3d$ $N=2$ $SO(24)$ Yang-Mills theory with some additional neutral
fields.
We also have no reason to expect this three-dimensional
theory to spontaneously
break supersymmetry \AHW.  Nevertheless \edthree, the massive states in
such a theory do not fall into supermultiplets and our
candidate duality between this theory and a nonsupersymmetric one
is possible.

\newsec{Conclusions}

It is interesting that we naturally find ourselves working
in 2+1 dimensions when we search for
candidate critical nonsupersymmetric string solitons in supersymmetric string
vacua.
In particular, it has recently been suggested that a nonsupersymmetric
world in four dimensions could be dual to a supersymmetric
world in three dimensions \edthree.  Such a
relation would provide a natural explanation
of the vanishing of the cosmological constant.
It is possible that our construction is providing a realization of such ideas,
but in order to verify this one would need a better understanding of the
worldsheet theories on the soliton string.
To begin with, one would need to gain better control over (1,1) conformally
invariant sigma models in general, as well as (0,2) models with
fractional R-charges.
Independent of any relation to
the ideas of \edthree, it seems to us that the nonsupersymmetric
string solitons discussed here should be
interpreted as evidence that there are indeed
nonsupersymmetric/supersymmetric dual pairs of string
vacua awaiting further exploration.  A related issue is the role
in supersymmetry breaking,
if any, of the plethora of {\it noncritical} nonsupersymmetric solitons
in string theory.

\bigskip
\centerline{\bf Acknowledgements}
\medskip
We would like to thank P. Aspinwall, C. Callan, P. Candelas,
B. Greene, C. Vafa, E. Witten, and
S-T Yau
for discussions.  S.K. is supported by the Harvard Society of Fellows and
by the William F. Milton Fund of Harvard University.
E.S. thanks the NSF and AT$\&$T GRPW for support.
\listrefs

\end